\documentclass[fleqn]{amsart}

\usepackage{amssymb}
\usepackage{amsfonts}
\usepackage[english]{babel}
\usepackage{euscript}
\usepackage{bbm}
\usepackage{xfrac}
\usepackage{color}
\usepackage{pb-diagram}

\theoremstyle{definition}

\theoremstyle{remark}

\numberwithin{equation}{section}

\newcommand{\fl}{\hspace*{-\mathindent}}
\newcommand{\textfrac}[2]{\textstyle{\frac{#1}{#2}}}
\newcommand{\W}{\EuScript{W}}

\begin{document}

\title[Lax representations for 2D Euler equation]{Extensions of the symmetry algebra and Lax representations for the two-dimensional
\\
 Euler equation}

\author{Oleg I. Morozov}

\address{Trapeznikov Institute of Control   Sciences, 65 Profsoyuznaya street, Moscow 117997, Russia}

\email{oimorozov@gmail.com}

\subjclass[2020]{58H05, 58J70, 35A30, 37K05, 37K10}

\date{\today  
}

\keywords{Two-dimensional Euler equation in the vorticity form;
symmetry algebra;
twisted extension;
Lie--Rinehart algebra;  
Lax representation;
non-removable parameter}

\begin{abstract}
We find  the twisted extensions of the symmetry algebra of the 2D Euler equation in the vorticity form and use them to construct new Lax repre\-sen\-ta\-ti\-on for this equation. Then we generalize this result by considering the transformation  Lie--Rinehart algebras generated by finite-dimensional subalgebras of the symmetry algebra and derive a family of Lax representations for the Euler equation. The family depends on functional parameters and con\-tains a non-removable spectral parameter.  
\end{abstract}

\maketitle


\begin{center} 
{\it Dedicated to Peter Olver on the occasion of his 70th birthday}.
\end{center}

\vskip 20 pt

\section{Introduction}

In this paper,  we consider 
the two dimensional Euler equation in  the vorticity form, 
\cite[\S 10]{LandauLifshits},
\begin{equation} 
\Delta (u_t) = [u, \Delta (u)], 
\label{Euler_eq}
\end{equation}
where $\Delta (u) = u_{xx}+u_{yy}$ and $[u, v] = u_x\,v_y-u_y\,v_x$.  
This equation is one of the   fundamental models of hydrodynamics  and the subject of intensive study, see \cite{ArnoldKhesin} and references therein. Our research is focused on the  Lax representations for equation \eqref{Euler_eq}. Lax representations provide the basic construction that allows ap\-pli\-ca\-ti\-ons of a num\-ber of techniques for studying nonlinear partial differential equations,  whence   they are considered as the key feature indicating integrability thereof, see
\cite{WE,Zakharov82,RogersShadwick1982,NovikovManakovPitaevskiyZakharov1984,%
Konopelchenko1987,AblowitzClarkson1991,MatveevSalle1991,Olver1993,BacklundDarboux2001}
and references therein. From the viewpoint of geometry of differential equations, Lax representations and other nonlocal structures of integrable nonlinear systems are naturally formulated in the language of differential coverings, 
\cite{KrasilshchikVinogradov1984,KrasilshchikVinogradov1989,VK1999}.

The Lax representation 
\begin{equation}
\left\{
\begin{array}{rcl}
p_t &=&   [u, p],
\\
\,
[\Delta (u), p] &=&  \lambda\,p
\end{array}
\right.
\label{Li_Lax_pair}
\end{equation} 
for equation \eqref{Euler_eq} was found in \cite{Li2001}, see also references therein and  \cite{LiShvidkoy2004}. 
The pa\-ra\-me\-ter $\lambda$ in system \eqref{Li_Lax_pair}  is removable. Indeed, when $\lambda \neq 0$,  the change of the 
pseu\-do\-po\-ten\-ti\-al $p =\mathrm{exp}(\lambda\,w)$ transforms  \eqref{Li_Lax_pair} to the form
\begin{equation}
\left\{
\begin{array}{rcl}
w_t &=&   [u, w],
\\
\,
[\Delta (u), w] &=&  1.
\end{array}
\right.
\label{Li_covering}
\end{equation} 
The Lax representation \eqref{Li_Lax_pair} with $\lambda = 0$  was used  in  \cite{LiYurov2003} to find a weak Darboux 
transformation for equation \eqref{Euler_eq}. Further generalizations of this Darboux transformation were presented and applied to find exact solutions for \eqref{Euler_eq} in \cite{LouLi2006,LouJiaTangHuang2007}. In \cite{LouJiaHuangTang2007}, the Lax representation \eqref{Li_Lax_pair} with $\lambda=0$ was used to define a special B{\"a}cklund transformation  for \eqref{Euler_eq}, and the transformation was utilized to obtain a number of exact solutions.  A dressing method for constructing solutions to equation \eqref{Euler_eq} was proposed in  \cite{YurovYurova2006}.

In the series of recent  papers \cite{Morozov2017} -- \cite{Morozov2021c} we developed the method for finding Lax representations of nonlinear {\sc pde}s. The method is based on the twis\-ted extensions of the Lie symmetry algebras of the {\sc pde}s under the study.  In \cite{Morozov2022} we  extend the technique to the Lie--Rinehart algebras, in particular, we have shown that extensions of Lie algebras by appending an integral of a non-trivial 1-cocycle  are useful in constructing Lax representations.      

In the present paper we apply the approach of \cite{Morozov2017} -- \cite{Morozov2022} to equation \eqref{Euler_eq}. We find the twisted extension of the symmetry algebra thereof. The linear combination \eqref{tau_1} of the Maurer--Cartan forms of the extension provides a Lax representation \eqref{first_covering_for_Euler_eq} without a non-removable parameter. This Lax representation differs from \eqref{Li_Lax_pair} with $\lambda=0$  and from  \eqref{Li_covering}.   

We employ a further generalization of the method of \cite{Morozov2017} -- \cite{Morozov2022} based of the definition of the transformation Lie--Rinehart algebras generated by finite-dimensional subalgebras exposed in \S 2.3.  Upon using this construction we derive a family of Lax representations for the Euler equation that depends on functional parameters and con\-tains a non-removable spectral parameter.  At the special choices of the parameters the obtained Lax representations get the form  \eqref{Li_Lax_pair} with 
$\lambda=0$  or \eqref{Li_covering}.


\section{Preliminaries and notation}

All the considerations in the paper are local.

\subsection{Symmetries and differential coverings}

The presentation in this subsection closely follows 
\cite{KrasilshchikVinogradov1984,KrasilshchikVinogradov1989,VK1999}.
Let $\pi \colon \mathbb{R}^n \times \mathbb{R}^m \rightarrow \mathbb{R}^n$,
$\pi \colon (x^1, \dots, x^n, u^1, \dots, u^m) \mapsto (x^1, \dots, x^n)$, be a trivial bundle, and
$J^\infty(\pi)$ be the bundle of its jets of infinite order. The local coordinates on $J^\infty(\pi)$ are
$(x^i,u^\alpha,u^\alpha_I)$, where $I=(i_1, \dots, i_n)$ are multi-indices, and for every local section
$f \colon \mathbb{R}^n \rightarrow \mathbb{R}^n \times \mathbb{R}^m$ of $\pi$ the corresponding infinite jet
$j_\infty(f)$ is a section $j_\infty(f) \colon \mathbb{R}^n \rightarrow J^\infty(\pi)$ such that
$u^\alpha_I(j_\infty(f))=\displaystyle{\frac{\partial ^{\#I} f^\alpha}{\partial x^I}}
=\displaystyle{\frac{\partial ^{i_1+\dots+i_n} f^\alpha}{(\partial x^1)^{i_1}\dots (\partial x^n)^{i_n}}}$.
We put $u^\alpha = u^\alpha_{(0,\dots,0)}$. Also, we will simplify notation in the following way: e.g., for $n=3$, $m=1$ 
we denote $x^1 = t$, $x^2= x$, $x^3= y$, and $u^1_{(i,j,k)}=u_{{t \dots t}{x \dots x}{y \dots y}}$ with $i$  times $t$,
$j$  times $x$, and $k$  times $y$.

The  vector fields
\[
D_{x^k} = \frac{\partial}{\partial x^k} + \sum \limits_{\# I \ge 0} \sum \limits_{\alpha = 1}^m
u^\alpha_{I+1_{k}}\,\frac{\partial}{\partial u^\alpha_I},
\qquad k \in \{1,\dots,n\},
\]
$(i_1,\dots, i_k,\dots, i_n)+1_k = (i_1,\dots, i_k+1,\dots, i_n)$,  are called {\it total derivatives}.
They com\-mu\-te everywhere on $J^\infty(\pi)$.

The {\it evolutionary vector field} associated to an arbitrary vector-valued smooth function
$\varphi \colon J^\infty(\pi) \rightarrow \mathbb{R}^m $ is the vector field
\[
\mathbf{E}_{\varphi} = \sum \limits_{\# I \ge 0} \sum \limits_{\alpha = 1}^m
D_I(\varphi^\alpha)\,\frac{\partial}{\partial u^\alpha_I}
\]
with $D_I=D_{(i_1,\dots\,i_n)} =D^{i_1}_{x^1} \circ \dots \circ D^{i_n}_{x^n}$.

A system of {\sc pde}s $F_r(x^i,u^\alpha_I) = 0$ of the order $s \ge 1$ with $\# I \le s$, $r \in \{1,\dots, R\}$
for some $R \ge 1$, defines the submanifold
$\EuScript{E}=\{(x^i,u^\alpha_I)\in J^\infty(\pi)\,\,\vert\,\,D_K(F_r(x^i,u^\alpha_I))=0,\,\,\# K\ge 0\}$
in $J^\infty(\pi)$.

A function $\varphi \colon J^\infty(\pi) \rightarrow \mathbb{R}^m$ is called a {\it (generator of an
infinitesimal) symmetry} of equation $\EuScript{E}$ when $\mathbf{E}_{\varphi}(F) = 0$ on $\EuScript{E}$. The
symmetry $\varphi$ is a solution to the {\it defining system}
\begin{equation}
\ell_{\EuScript{E}}(\varphi) = 0
\label{defining_eqns}
\end{equation}
of equation $\EuScript{E}$, where $\ell_{\EuScript{E}} = \ell_F \vert_{\EuScript{E}}$ with the matrix differential operator
\[
\ell_F = \left(\sum \limits_{\# I \ge 0}\frac{\partial F_r}{\partial u^\alpha_I}\,D_I\right).
\]
The {\it symmetry algebra} $\mathrm{Sym} (\EuScript{E})$ of equation $\EuScript{E}$ is the linear space of
solutions to  (\ref{defining_eqns}) endowed with the structure of a Lie algebra over $\mathbb{R}$ by the
{\it Jacobi bracket} $\{\varphi,\psi\} = \mathbf{E}_{\varphi}(\psi) - \mathbf{E}_{\psi}(\varphi)$.
The {\it algebra of contact symmetries} $\mathrm{Sym}_0 (\EuScript{E})$ is the Lie subalgebra of
$\mathrm{Sym} (\EuScript{E})$ defined as $\mathrm{Sym} (\EuScript{E}) \cap C^\infty(J^1(\pi))$.

Let the linear space $\EuScript{V}$ be either $\mathbb{R}^N$ for some $N \ge 1$ or  $\mathbb{R}^\infty$
endowed with  local co\-or\-di\-na\-tes $v^a$, $a \in \{1, \dots , N\}$ or  $a \in  \mathbb{N}$, respectively.
The variables $v^a$ are cal\-led {\it pseudopotentials} \cite{WE}.  
Locally, a {\it differential covering} of $\EuScript{E}$ is 
a trivial bundle $\varpi \colon J^\infty(\pi) \times \EuScript{V} \rightarrow J^\infty(\pi)$ equipped with the {\it extended total derivatives}
\[
\widetilde{D}_{x^k} = D_{x^k} + \sum \limits_{ s =0}^\infty
T^s_k(x^i,u^\alpha_I,v^j)\,\frac{\partial }{\partial v^s}
\]
such that $[\widetilde{D}_{x^i}, \widetilde{D}_{x^j}]=0$ for all $i \not = j$ on $ \EuScript{E}$.
Define the par\-ti\-al derivatives of $v^s$ by  $v^s_{x^k} =  \widetilde{D}_{x^k}(v^s)$.  This gives the over-determined system
of {\sc pde}s
\begin{equation}
v^s_{x^k} = T^s_k(x^i,u^\alpha_I,v^j)
\label{WE_prolongation_eqns}
\end{equation}
which is compatible whenever $(x^i,u^\alpha_I) \in \EuScript{E}$.
System \eqref{WE_prolongation_eqns} is referred to as the {\it covering equations} or the {\it Lax representation} of 
equation $\EuScript{E}$. 

 Dually, the differential covering is defined by the
{\it Wahlquist--Estabrook forms}
\begin{equation}
\tau^s = d v^s - \sum \limits_{k=1}^{n} T^s_k(x^i,u^\alpha_I,v^j)\,dx^k
\label{WEfs}
\end{equation}
as follows: when $v^s$  and $u^\alpha$ are considered to be functions of $x^1$, ... , $x^n$, forms \eqref{WEfs}
are equal to zero whenever  system \eqref{WE_prolongation_eqns} holds.

Two differential coverings $\varpi_1$ and $\varpi_2$ of equation $\EuScript{E}$ 
with the extended total derivatives $\widetilde{D}_{x^k}^{(1)}$ 
and $\widetilde{D}_{x^k}^{(2)}$ 
are called {\it equivalent} if  there exists a diffeomorphism $\Phi$ such that the diagram  
\[
\begin{diagram}
\node{\EuScript{E}\times \EuScript{V}_1}
     \arrow[2]{e,t}{\Phi}
     \arrow{se,r}{\varpi_1\vert_{\EuScript{E}}}
\node[2]{\EuScript{E}\times \EuScript{V}_2}
      \arrow{sw,r}{\varpi_2\vert_{\EuScript{E}}}                            
\\
\node[2]{\EuScript{E}
}
\end{diagram}
\]
is commutative and $\Phi_{*}(\widetilde{D}_{x^k}^{(1)})=\widetilde{D}_{x^k}^{(2)}$.
In other words, if $\tau^{(1),s}$ and $\tau^{(2),s}$ are the Wahlquist--Estabrook forms of $\varpi_1$ and $\varpi_2$, 
then the coverings are equivalent when
\begin{equation}
\Phi^{*} \tau^{(2),s} \in 
\langle\,
\tau^{(1),m}, \vartheta^\alpha_I\vert_{\EuScript{E}} \,\,\vert\,\, m \in \mathbb{N}, \,\alpha \in \{1,...,m\}, \,\# I \ge 0\,\rangle,
\label{equivalence_of_coverings}
\end{equation}
where $\vartheta^\alpha_I = du^\alpha_I- \sum \limits_{k=1}^{n} u^\alpha_{I+i_k} dx^k$  are the contact forms on 
$J^\infty(\pi)$.


\subsection{Twisted extensions of Lie algebras}

Consider a Lie algebra $\mathfrak{g}$ over $\mathbb{R}$ with the non-trivial first cohomology group $H^1(\mathfrak{g})$ and take a non-trivial closed 1-form $\alpha$ on $\mathfrak{g}$. Then for any $c \in \mathbb{R}$ define new differential
$d_{c \alpha} \colon C^k(\mathfrak{g},\mathbb{R}) \rightarrow C^{k+1}(\mathfrak{g},\mathbb{R})$ by the formula 
$d_{c \alpha} \theta = d \theta - c \,\alpha \wedge \theta$. From  $d\alpha = 0$ it follows that $d_{c \alpha} ^2=0$. The cohomology groups of the complex
\[
C^1(\mathfrak{g}, \mathbb{R})
\stackrel{d_{c \alpha}}{\longrightarrow}
\dots
\stackrel{d_{c \alpha}}{\longrightarrow}
C^k(\mathfrak{g}, \mathbb{R})
\stackrel{d_{c \alpha}}{\longrightarrow}
C^{k+1}(\mathfrak{g}, \mathbb{R})
\stackrel{d_{c \alpha}}{\longrightarrow} \dots
\]
are referred to as the {\it twisted} {\it cohomology groups}  \cite{Novikov2002,Novikov2005} of $\mathfrak{g}$
and denoted by $H^{*}_{c\alpha}(\mathfrak{g})$.

Suppose that for a Lie algebra $\mathfrak{g}$ with the Maurer--Cartan forms  $\{\omega^i \,\,\vert\,\, i \in \mathbb{N}\}$ 
and the structure equations 
\begin{equation}
d\omega^i = \sum \limits_{j<k}\,a^i_{jk} \,\omega^j \wedge \omega^k 
\label{g_se} 
\end{equation}
there hold $H^1(\mathfrak{g}) \neq \{0\}$  and there is $c_0 \in \mathbb{R}$ such that 
$H^2_{c_0\alpha}(\mathfrak{g}) \neq\{[0]\}$. Then for a non-trivial twisted 2-cocycle $\Omega$ such that   
$[\Omega] \in H^2_{c_0\alpha}(\mathfrak{g}) \setminus  \{[0]\}$ the equation
\begin{equation}
d\sigma = c_0\,\alpha\wedge \sigma+\Omega
\label{extension_se}
\end{equation}    is compatible with equations \eqref{g_se}. The Lie algebra $\hat{\mathfrak{g}}$ with the Maurer--Cartan forms 
$\{\omega^i, \sigma \}$ and the structure equations \eqref{g_se}, \eqref{extension_se}  is referred to as the 
{\it twisted extension of} $\mathfrak{g}$ {\it generated by the cocycle} $\Omega$.

\subsection{Transformation Lie--Rinehart algebras}
Let  $\mathfrak{q}$ be a Lie algebra  and 
$\varrho$ be an infinitesimal action of  $\mathfrak{q}$ on an open subset $\W \subseteq \mathbb{R}^N$ for some 
$N \in\mathbb{N}$, that is, a representation
$\varrho \in \mathrm{Hom}_{\mathbb{R}} (\mathfrak{q}, \mathrm{Der}\,(C^\infty(\W)))$. 
Consider the vector space $\mathfrak{q}_{C^\infty(\W)} = C^\infty(\W) \otimes \mathfrak{q}$ equipped with a structure of a left $C^\infty(\W)$-module and with the map
$\mathrm{id}_{C^\infty(\W)} \otimes \varrho \colon C^\infty(\W) \otimes \mathfrak{q} \rightarrow \mathrm{Der}\,(C^\infty(\W))$,
which is a homomorphism of Lie algebras over $\mathbb{R}$ and a homomorphism  of $C^\infty(\W)$-modules.  Then 
$\mathfrak{q}_{C^\infty(\W)}$ is referred to as a {\it transformation Lie--Rinehart algebra}  
 associated with $\mathfrak{q}$ and $\varrho$ provided there holds\footnote{We write 
$a\,x$ instead of $a\otimes x$ for $a \in   C^\infty(\W)$,  $x \in \mathfrak{q}$.}
\[
{}[x, a\,y] = a\,[x, y] +\varrho(x)(a)\,y
\] 
for $x, y \in  \mathfrak{q}_{C^\infty(\W)}$ and $a \in C^\infty(\W)$, see  \cite{Mackenzie1995}.   Let $\{x_m\,\,\vert\,\,m \ge 1\}$ be a basis of $\mathfrak{q}$ with the structure constants $c^i_{jk}$ defined by $[x_j, x_k] = - \sum_{i}  c^i_{jk}\, x_i$  and let 
$w =(w_1, .... ,w_N)$ be local coordinates on $\W$. The representation $\varrho$ is defined  by the equations 
\[
\varrho (x_m) = \sum\limits_{i=1}^N h_m^i(w)\,\partial_{w_i}  
\]   
for a collection of functions $h_m^i \in C^\infty(\W)$.
Let $\{\beta^m \,\,\vert\,\,m \ge 1\}$  be the Maurer--Cartan forms of $\mathfrak{q}$  that are dual to the basis, 
that is, $\beta^m(x_n)=\delta^m_n$.  The structure equations of $\mathfrak{q}$ have the form
\begin{equation}
d\beta^i_{jk} =\sum \limits_{j<k} \,c^i_{jk} \,\beta^j \wedge \beta^k.
\label{SE_of_q}
\end{equation} 
Then   $\mathfrak{q}_{C^\infty(\W)} = \langle g(w)\,x_k \,\,\vert \,\, k \ge 1, g \in C^{\infty}(\W) \rangle$,
and the structure equations for $\mathfrak{q}_{C^\infty(\W)}$ are obtained by appending the equations 
\begin{equation}
dw^i = \sum\limits_{m \ge 1} h_m^i\,\beta^m
\label{dws}
\end{equation}
to equations \eqref{SE_of_q}. The Jacobi identity for elements of $\mathfrak{q}_{C^\infty(\W)}$
imply that the conditions  $d(dw^i) =0$  are consequences of equations \eqref{SE_of_q} and \eqref{dws}. 

The symmetry algebras of integrable {\sc pde}s often have the structure of a semi-direct sum  
$\mathfrak{q} = \mathfrak{q}_\diamond \ltimes \mathfrak{q}_\infty$ of a finite-dimensional Lie algebra $\mathfrak{q}_\diamond$ 
and an infinite-dimensional ideal $\mathfrak{q}_\infty$, see, e.g., \cite{KruglikovMorozov2015,Morozov2019}. We apply the above construction to a representation 
$\varrho \in \mathrm{Hom}_{\mathbb{R}}(\mathfrak{q},\mathrm{Der} (C^\infty(\W)))$ such that  
$\varrho \vert_{\mathfrak{q}_\infty} = 0$.  If $\mathrm{dim} \, \mathfrak{q}_\diamond=n$, 
$\mathfrak{q}_\diamond^{*} = \langle \beta_1, ... , \beta_n \rangle$, and
$\mathfrak{q}_\infty ^{*}= \langle \gamma_k \,\,\vert\,\,k \in \mathbb{N}\rangle$,
then we will refer the Lie--Rinehart algebra $\mathfrak{q}_{C^\infty(\W)}$ with 
$(\mathfrak{q}_{C^\infty(\W)})^{*} = \langle g(w)\,\beta_i, h(w)\,\gamma_k \,\,
\vert\,\, i \in \{1, ..., n\}, k \in \mathbb{N}, g, h \in C^\infty(\W) \rangle$ 
to as the {\it transformation Lie--Rinehart algebra generated by the subalgebra $\mathfrak{q}_\diamond$ and the 
representation $\varrho$}. The structure equations of $\mathfrak{q}_{C^\infty(\W)}$ are obtained by attaching the structure equations of the ideal $\mathfrak{q}_\infty$ to the structure equations \eqref{SE_of_q}, 
\eqref{dws} with with $i,j,k, m \in \{1, ... , n\}$.

In the particular case that $\mathrm{dim} \, \mathfrak{q}_\diamond=\mathrm{dim}\, \W = 1$, 
$\varrho (\mathfrak{q}_\diamond) =  \mathbb{R}\,\partial_w$,  $\mathfrak{q}_\diamond^{*} = \langle \beta \rangle$ with 
$d\beta = 0$,  the Lie--Rinehart algebra generated by $\mathfrak{q}_\diamond$ and $\varrho$ coincides with the 
Lie--Rinehart algebra obtained via procedure of extension of $\mathfrak{q}$  by appending an integral of 1-cocycle $\beta$ discussed in \cite[Remark 2]{Morozov2022}.


\section{The symmetry algebra and  the twisted extensions}

\subsection{The change of coordinates}

To simplify computations in the subsequent sections we perform the following change of variables: we write equation
\eqref{Euler_eq}  as   
$\tilde{u}_{\tilde{t}\tilde{x}\tilde{x}}+\tilde{u}_{\tilde{t}\tilde{y}\tilde{y}} 
= 
\tilde{u}_{\tilde{x}}\,(\tilde{u}_{\tilde{x}\tilde{x}\tilde{y}}+\tilde{u}_{\tilde{y}\tilde{y}\tilde{y}})
-\tilde{u}_{\tilde{y}}\,(\tilde{u}_{\tilde{x}\tilde{x}\tilde{x}}+\tilde{u}_{\tilde{x}\tilde{y}\tilde{y}})$
and then put 
$\tilde{t}=t$, 
$\tilde{x}=\frac{1}{2}\,(1+\mathrm{i})\,(x+y)$, 
$\tilde{y} =-\frac{1}{2}\,(1-\mathrm{i})\,(x-y)$,  
and $\tilde{u}=u$, where $\mathrm{i} =\sqrt{-1}$. 
This yields the transformed Euler equation
\begin{equation}
u_{txy} = u_x\,u_{xyy}-u_y\,u_{xxy},
\label{transformed_Euler_eq}
\end{equation}
We can write this equation in the form similar to \eqref{Euler_eq} as    
$\mathrm{D}(u_t) = [u, \mathrm{D}(u)]$ with  $\mathrm{D}(u) = u_{xy}$.

\subsection{The generators, the Maurer--Cartan forms, and the structure equations of the contact symmetry  algebra}

The Lie algebra $\mathrm{Sym}_0 (\EuScript{E})$ of the contact infinitesimal symmetries of equation \eqref{transformed_Euler_eq} has generators 
\[
\begin{array}{rcl}
\psi_0 &=& -t\,u_t -u,
\\
\psi_1 &=& -u_t,
\\
\psi_2 &=& t\,x\,u_x-t\,y\,u_y-x\,y,
\\
\psi_3 &=& t\,u_t-x\,u_x+ y\,u_y+u,
\\
\psi_4 &=& t\,u_t-y\,u_y+2\,u,
\\
\varphi_1(A_1) &=& -A_1\,u_x+A_1^{\prime}\,y,
\\
\varphi_2(A_2) &=& -A_2\,u_y-A_2^{\prime}\,x,
\\
\varphi_3(A_3) &=& A_3,
\end{array}
\]
where $A_i=A_i(t)$  are arbitrary smooth functions of $t$.   For the purposes of the present study we restrict the attention to the polynomial functions $A_i$, that is,  we consider the subalgebra $\mathfrak{s} \subset \mathrm{Sym}_0 (\EuScript{E})$ generated by $\psi_i$ and $\varphi_j(t^k)$  with  $i \in \{0, ..., 4\}$, $j \in\{1,2,3\}$, and  $k \in \mathbb{N} \cup \{ 0 \}$. Then the 
Maurer--Cartan forms $\alpha_i$, $\omega_{jk}$ of $\mathfrak{s}$ can be defined  as the dual 1-forms to the generators, that is,  by imposing the equations
$\alpha_i(\psi_{i^{\prime}}) = \delta_{ii^{\prime}}$,  $\alpha_i(\varphi_j(t^k))=0$, 
$\omega_{jk}(\psi_i) = 0$,    and 
$\omega_{jk}(\varphi_{j^\prime}(t^{k^\prime})) = \delta_{jj^{\prime}}\,\delta_{kk^\prime}$. 
Using the technique of moving frames \cite{Olver_Pohjanpelto_2005,Cheh_Olver_Pohjanpelto_2005,Olver_Pohjanpelto_Valiquette_2009}, we can write 
the structure equations of $\mathfrak{s}$ in the form
\begin{equation}
\left\{
\begin{array}{rcl}
d\alpha_0 &=& 0,
\\
d\alpha_1 &=& \alpha_0 \wedge \alpha_1,
\\
d\alpha_2 &=& -\alpha_0 \wedge \alpha_2,
\\
d\alpha_3 &=& \alpha_1 \wedge \alpha_2,
\\
d\alpha_4 &=& 0,
\\
d\Omega_1 &=& \Omega_{1,h} \wedge (h\,\alpha_0 +\alpha_1) +\Omega_1 \wedge (h\,\alpha_2+\alpha_0-\alpha_3 ),
\\
d\Omega_2 &=& \Omega_{2,h} \wedge (h\,\alpha_0 +\alpha_1) +\Omega_2 \wedge (h\,\alpha_2-\alpha_0-\alpha_3+\alpha_4),
\\
d\Omega_3 &=& \Omega_{3,h} \wedge (h\,\alpha_0 +\alpha_1)
+ (\alpha_4-3\,\alpha_0)\wedge \Omega_3  +\Omega_{1,h} \wedge \Omega_2 
\\
&&
+ \Omega_{1} \wedge \Omega_{2,h},
\end{array}
\right.
\label{SE}
\end{equation}
where 
\[
\Omega_j = \sum \limits_{k=0}^{\infty} \frac{h^k}{k!}\,\omega_{jk}
\]
are the formal series with the formal parameter $h$ such that $dh=0$ and $\Omega_{j,h} = \partial_h \Omega_j$ are the formal derivatives of $\Omega_j$ with respect to $h$. The Lie algebra $\mathfrak{s}$ is not of the Kac--Moody type, see 
\cite[\S~3]{Morozov2021c}.  We can find the Maurer--Cartan forms by integrating the structure equations and by implementing 
Cartan's equivalence method, \cite{Olver1995,FelsOlver1998,Morozov2002}. In the subsequent sections we need the explicit expressions for the following forms:
\[
\alpha_0 =\frac{dq}{q}, \quad \alpha_1 = q\,dt, \quad \alpha_2 = \frac{du_{xy}}{q},
\quad 
\alpha_3 = \frac{du_{xyy}}{u_{xyy}}-u_{xy}\,dt,
\]
\[
\alpha_{4} = \frac{du_{xxy}}{u_{xxy}} + \frac{du_{xyy}}{u_{xyy}},
\quad
\omega_{10} = \frac{u_{xyy}}{q}\,(dy +u_x\,dt), 
\quad
\omega_{20} = \frac{u_{xxy}}{q}\,(dx -u_y\,dt), 
\]
\[
\omega_{30} = \frac{u_{xxy}u_{xyy}}{q^3}\,(du-u_t\,dt-u_x\,dx-u_y\,dy),
\]
In these forms $q$ is a non-zero parameter.

\subsection{The twisted extension}

From the structure equations \eqref{SE} it follows that  $H^1(\mathfrak{s}) = \langle \alpha_0, \alpha_4\rangle$.
The vector field $U = t\,\partial_t+x\,\partial_x+y\,\partial_y+u\,\partial_u$ associated to the symmetry 
$4\,\psi_0+\psi_3+2\,\psi_4 = -t\,u_t-x\,u_x-y\,u_y+u$ defines an inner grading, \cite[\S 1.5.2]{Fuks1984}, on the Lie algebra
$\mathfrak{s}$. The computations similar to those used in the proof of Theorem 2 from \cite{Morozov2018} give
\[
H^2_{c_1  \alpha_0+c_2 \alpha_4}(\mathfrak{s}) =
\left\{
\begin{array}{lcl}
\langle [\alpha_1 \wedge \omega_{30}+\omega_{10}\wedge\omega_{20}]\rangle,
&~~~& c_1 =-2,  \,\, c_2 = 1,
\\
\langle [\alpha_{0}\wedge \alpha_2],   [\alpha_2 \wedge \alpha_3], [\alpha_2\wedge \alpha_4] \rangle, & & c_1 = -1, \,\,c_2 =0, 
\\
\langle [\alpha_{0}\wedge \alpha_1],   [\alpha_1 \wedge \alpha_3], [\alpha_1\wedge \alpha_4] \rangle, & & c_1 = 1, \,\,c_2 =0,
\\
\langle [\alpha_{0}\wedge \alpha_4] \rangle, & & c_1 = 0, \,\,c_2 =0,
\\
\{[0]\}, && \mathrm{otherwise}. 
\end{array}
\right.
\]
The non-trivial twisted 2-cocycles define the eight-dimensional twisted extension 
$\widehat{\mathfrak{s}}$
of the Lie algebra $\mathfrak{s}$. The structure equations of  $\widehat{\mathfrak{s}}$ include 
equations \eqref{SE} and the system  
\begin{equation}
\left\{
\begin{array}{lcl}
d\sigma_1 &=& (\alpha_4-2\,\alpha_0) \wedge \sigma_1 +\alpha_1 \wedge \omega_{30}+\omega_{10}\wedge\omega_{20},
\\
d\sigma_2 &=& -\alpha_0 \wedge \sigma_2+ \alpha_{0}\wedge \alpha_2,
\\
d\sigma_3 &=& -\alpha_0 \wedge \sigma_3+ \alpha_{2}\wedge \alpha_3,
\\
d\sigma_4 &=& -\alpha_0 \wedge \sigma_4+ \alpha_{2}\wedge \alpha_4,
\\
d\sigma_5 &=& \alpha_0 \wedge \sigma_5+ \alpha_{0}\wedge \alpha_1,
\\
d\sigma_6 &=& \alpha_0 \wedge \sigma_6+ \alpha_{1}\wedge \alpha_3,
\\
d\sigma_7 &=& \alpha_0 \wedge \sigma_7+ \alpha_{1}\wedge \alpha_4,
\\
d\sigma_8 &=& \alpha_0 \wedge \alpha_4.
\end{array}
\right.
\label{additional_SE}
\end{equation} 
In what follows we need the explicit expression for the 1-form 
\begin{equation}
\sigma_1 = \frac{u_{xxy}u_{xyy}}{q^2}(dv-u\,dt +\textfrac{1}{2}\,(y\,dx-x\,dy))
\label{sigma_1}
\end{equation}
obtained  by integration of the first equation in \eqref{additional_SE} with the known 1-forms $\alpha_0$, $\alpha_4$, and 
$\omega_{j0}$.  The new variable $v$ in \eqref{sigma_1} is the `constant of integration'.


\section{Lax representations}

\subsection{The Wahlquist--Estabrook forms and the Lax representations}

To find a Lax representation for equation \eqref{transformed_Euler_eq}
we consider the linear combination
\begin{equation}
\tau_1 = \sigma_1 - \omega_{10}-\omega_{20}
\label{tau_1}
\end{equation}  
\[
=\frac{u_{xxy}u_{xyy}}{q^2}\,\left(
dv - \frac{1}{2\,u_{xyy}}\,(2\,q-y\,u_{xyy})\,dx-  \frac{1}{2\,u_{xxy}}\,(2\,q+x\,u_{xyy})\,dy
\right.
\]
\[
\qquad\qquad\qquad\qquad\qquad\qquad
\left.
-\left(\frac{u_xu_{xyy}-u_yu_{xxy}}{u_{xxy}u_{xyy}}\,q+u\right)\,dt
\right).
\]
We rename
\[
q = (v_x+\textfrac{1}{2}y)\,u_{xyy}
\]
to make the coefficient at $dx$ inside the outer parentheses equal to $v_x$. This yields the Wahlquist--Estabrook form 
\[  
\fl
\tau_1= \frac{u_{xxy}}{(v_x+\textfrac{1}{2}y)^2\,u_{xyy}}
\left(
dv -v_x\,dx 
-\left(\frac{u_x\,u_{xyy}-u_y\,u_{xxy}}{u_{xxy}}\,\left(v_x+\textfrac{1}{2}y\right) +u\right)\,dt
\right.
\]
\[
\qquad\qquad\qquad\qquad\qquad\qquad
\left.
-\frac{1}{u_{xxy}}\,\left(u_{xyy}\,v_x+\frac{1}{2}\,(x\,u_{xxy}+y\,u_{xyy})\right)\,dy
\right)
\]
of the Lax representation 
\[
\left\{
\begin{array}{lcl}
v_t &=& \displaystyle{\frac{u_x\,u_{xyy}-u_y\,u_{xxy}}{u_{xxy}}\,\left(v_x+\textfrac{1}{2}y\right)+u},
\\
v_y &=& \displaystyle{\frac{1}{u_{xxy}} \,\left(u_{xyy}\,v_x+\frac{1}{2}\,(x\,u_{xxy}+y\,u_{xyy})\right)}
\end{array}
\right.
\]
for equation \eqref{transformed_Euler_eq}. We can write this system in the form
\begin{equation}
\left\{
\begin{array}{rcl}
\,v_t &=& [u,v]+u-\textfrac{1}{2}\,\mathrm{E}(u), 
 \\
\,[\mathrm{D}(u), v] &=& \textfrac{1}{2} \,\mathrm{E}(\mathrm{D}(u)), 
\end{array}
\right.
\label{first_covering}
\end{equation}
where $\mathrm{E}(u) = x\,u_x+y\,u_y$.   Direct computations show that the compatibility conditions for system \eqref{first_covering}  follow from equation \eqref{transformed_Euler_eq}. 
Using the transformation inverse to the change of variables in \S 3.1 we get the Lax  representation  
\begin{equation}
\left\{
\begin{array}{rcl}
\,v_t &=& [u,v]+u-\textfrac{1}{2}\,\mathrm{E}(u), 
 \\
\,[\Delta(u), v] &=& \textfrac{1}{2} \,\mathrm{E}(\Delta(u)) 
\end{array}
\right.
\label{first_covering_for_Euler_eq}
\end{equation}
for the Euler equation. This Lax representation differs from \eqref{Li_Lax_pair} with $\lambda =0$ and from \eqref{Li_covering}.

To find  other Lax representations for equation \eqref{transformed_Euler_eq} we will use the transformation Lie--Rinehart  algebras 
generated by finite-dimensional subalgebras of the Lie algebra $\mathfrak{s}$. As the first example we consider the Lie--Rinehart algebra $\widehat{\mathfrak{s}}_{C^\infty(\mathbb{R})}$ generated by the subalgebra $\langle \psi_0 \rangle$ of the Lie  algebra 
$\widehat{\mathfrak{s}}$ constructed in \S 3.3 and the representation $\varrho_1$ of  $\widehat{\mathfrak{s}}$ such that 
$\varrho_1(\psi_0) = q\,\partial_q$ and $\varrho_1(\psi_i)=\varrho_1(\varphi_k(t^m)) = \varrho_1(\sigma_j) = 0$
for $i \in \{1,.., 4\}$, $k \in \{1,2,3\}$, $m \ge 0$, and  $j \in \{1, ... ,8\}$. The Maurer--Cartan forms of 
$\widehat{\mathfrak{s}}_{C^\infty(\mathbb{R})}$ are obtained via multiplying the Maurer--Cartan forms of the Lie algebra 
$\widehat{\mathfrak{s}}$ by functions of $q$, the structure equations of $\widehat{\mathfrak{s}}_{C^\infty(\mathbb{R})}$
are the collection of the structure equations \eqref{SE}, \eqref{additional_SE}, and the equation $dq = q\,\alpha_0$.
We take the linear combination 
\[
\tau_2= \sigma_1-\omega_{10} 
-\left(1-\frac{1}{q}\right)\, \omega_{20}   
\in 
\left(\widehat{\mathfrak{s}}_{C^\infty(\mathbb{R})}\right)^{*},
\] 
put
\[
q =1+ (v_x+\textfrac{1}{2}y)\,u_{xyy},
\label{q_substitution}
\]
and obtain  
\begin{equation}  
\tau_2= \frac{u_{xxy}u_{xyy}}{(1+ (v_x+\textfrac{1}{2}y)\,u_{xyy})^2}
\left(
dv -v_x\,dx      \phantom{\frac{u_x\,u_{xyy}-u_y\,u_{xxy}}{u_{xxy}}}
\right.
\label{def_tau_2}
\end{equation}
\[
\qquad\qquad
-\left(\frac{u_x\,u_{xyy}-u_y\,u_{xxy}}{u_{xxy}}\,\left(v_x+\textfrac{1}{2}y\right) +u+\frac{u_x}{u_{xxy}}\right)\,dt
\]
\[
\qquad\qquad
\left.
-\frac{1}{u_{xxy}}\,\left(u_{xyy}\,v_x+\frac{1}{2}\,(x\,u_{xxy}+y\,u_{xyy})+1\right)\,dy
\right).
\]
This Wahlquist--Estabrook form defines the Lax representation 
\[
\left\{
\begin{array}{lcl}
v_t &=& \displaystyle{\frac{u_x\,u_{xyy}-u_y\,u_{xxy}}{u_{xxy}}\,\left(v_x+\textfrac{1}{2}y\right)+u+\frac{u_x}{u_{xxy}}},
\\
v_y &=& \displaystyle{\frac{1}{u_{xxy}} \,\left(u_{xyy}\,v_x+\frac{1}{2}\,(x\,u_{xxy}+y\,u_{xyy})+1\right),}
\end{array}
\right.
\]
or
\begin{equation}
\left\{
\begin{array}{rcl}
\,v_t &=& [u,v]+u-\textfrac{1}{2}\,\mathrm{E}(u),      
\\
\,[\mathrm{D}(u), v] &=& 1+\textfrac{1}{2} \,\mathrm{E}(\mathrm{D}(u)).  
\end{array}
\right.
\label{second_covering}
\end{equation}
The compatibility conditions of this system are consequences of equation \eqref{transformed_Euler_eq}.

Furthermore, we take the subalgebra 
$\langle \psi_0, \psi_1, \psi_2, \psi_4 \rangle  \subset  \widehat{\mathfrak{s}}$  and the representation $\varrho_2$ of  the Lie algebra $\widehat{\mathfrak{s}}$ by vector fields on the space $\mathbb{R}^4$ with local coordinates  
$(w_1,w_2,w_3,w_4) = (q, t, u_{xy}, u_{xxy}u_{xyy})$ defined by the formulas 
$\varrho_2(\psi_0) = w_1\,\partial_{w_1}$, 
$\varrho_2(\psi_1) = w_1^{-1}\,\partial_{w_2}$,
$\varrho_2(\psi_2) = w_1\,\partial_{w_3}$,
$\varrho_2(\psi_4) = w_4\,\partial_{w_4}$,
$\varrho_2(\psi_3)=\varrho_2(\varphi_k(t^m)) = \varrho_2(\sigma_j) = 0$
for  $k \in \{1,2,3\}$, $m \ge 0$, and  $j \in \{1, ... ,8\}$. 
The Maurer--Cartan forms of the transformation Lie--Rinehart algebra $\widehat{\mathfrak{s}}_{C^\infty(\mathbb{R}^4)}$ are obtained via multiplying the Maurer--Cartan forms of the Lie algebra $\widehat{\mathfrak{s}}$ by functions of $w_1$, ... , $w_4$,  the structure equations of $\widehat{\mathfrak{s}}_{C^\infty(\mathbb{R}^4)}$ include the structure equations \eqref{SE}, \eqref{additional_SE}, and the equations $dw_1 = w_1\,\alpha_0$, $dw_2 = w_1^{-1}\,\alpha_1$, $dw_3 = w_1\,\alpha_2$, $dw_4=w_4\,\alpha_4$. We consider the linear combination 
\[
\tau_3 = \sigma_1 - \frac{w_1+G(w_3)}{w_1} \,\omega_{10} - \omega_{20}
-\frac{w_4\,F(w_3)}{w_1^3}\,\alpha_1
\]
of the Maurer--Cartan forms of $\widehat{\mathfrak{s}}_{C^\infty(\mathbb{R}^4)}$ with arbitrary smooth functions 
$F$ and $G$ of $w_3 = u_{xy}=\mathrm{D}(u)$. After substituting $q=u_{xyy}\,\left(v_x+\frac{1}{2}\,y\right)$
the form $\tau_3$ becomes the Wahlquist--Estabrook form of  the Lax representation
\begin{equation}
\left\{
\begin{array}{rcl}
\,v_t &=& [u,v]+u-\textfrac{1}{2}\,\mathrm{E}(u)+F(\mathrm{D}(u)),      
\\
\,[\mathrm{D}(u), v] &=& \textfrac{1}{2} \,\mathrm{E}(\mathrm{D}(u))+G(\mathrm{D}(u))  
\end{array}
\right.
\label{third_covering}
\end{equation}
of equation \eqref{transformed_Euler_eq}.  Systems 
\eqref{first_covering} and 
\eqref{second_covering} are particular cases of system \eqref{third_covering} 
that correspond to $F\equiv G \equiv 0$  and $F\equiv 0$, $G \equiv 1$, respectively. 


\subsection{The non-removable parameters}

The symmetry  $\psi_0 = -t\,u_t-u$ of equation \eqref{transformed_Euler_eq} does not admit a lift to a symmetry of equations \eqref{third_covering}  when 
\begin{equation}
\frac{G(w_3)}{w_3} \neq \mathrm{const}.
\label{G_condition}
\end{equation}
For the vector field $V_0 = t\,\partial_t - u\,\partial_u$ associated to $\psi_0$ we consider the  diffeomorphism
$\mathrm{exp}(\varepsilon \,\mathrm{pr}_3 V_0) \colon J^3(\pi) \rightarrow J^3(\pi)$ and its extension  
$\mathrm{exp}(\varepsilon \,\mathrm{pr}_3 V_0) \times \mathrm{id}  \colon J^3(\pi) \times \EuScript{V}\rightarrow J^3(\pi) \times \EuScript{V}$, where $\EuScript{V}$ is the domain of the pseudopotential  $v$. The pull-back of the prolongation of the resulting diffeomorphism maps the form $\tau_3$ to the Wahlquist--Estabrook form of the Lax representation 
\begin{equation}
\left\{
\begin{array}{rcl}
\,v_t &=& [u,v]+u-\textfrac{1}{2}\,\mathrm{E}(u)+\lambda\, F(\lambda^{-1}\, \mathrm{D}(u)),      
\\
\,[\mathrm{D}(u), v] &=& \textfrac{1}{2} \,\mathrm{E}(\mathrm{D}(u))
+\lambda\, G(\lambda^{-1}\, \mathrm{D}(u))  
\end{array}
\right.
\label{third_covering_lambda}
\end{equation} 
with $\lambda=\mathrm{e}^{\varepsilon}$.
In accordance with \cite[\S\S~3.2, 3.6]{KrasilshchikVinogradov1989}, cf.
\cite{Krasilshchik2000,IgoninKrasilshchik2000,Marvan2002,IgoninKerstenKrasilshchik2002},  this parameter 
is non-removable, that is, differential coverings defined by system \eqref{third_covering_lambda} with different constant values of 
$\lambda$ are not equivalent.  In particular, system \eqref{second_covering} gets the form
\begin{equation}
\left\{
\begin{array}{rcl}
\,v_t &=& [u,v]+u-\textfrac{1}{2}\,\mathrm{E}(u),  
\\
\,[\mathrm{D}(u), v] &=& \lambda +\textfrac{1}{2} \,\mathrm{E}(\mathrm{D}(u))
\end{array}
\right.
\label{second_covering_epsilon}
\end{equation}
with the  non-removable parameter $\lambda$.

When $G(w_3)=\gamma\,w_3$ with $\gamma \in \mathbb{R}\setminus \{ 0\}$, the symmetry  $\psi_2 =t\,x\,u_x-t\,y\,u_y-x\,y$ does not possess a lift to a symmetry of system \eqref{third_covering}.  Hence for the vector field 
$V_2= -t\,x\,\partial_x+t\,y\,\partial_y-x\,y\,\partial_u$ associated to $\psi_2$  the pull-back of  the prolongation of the 
diffeomorphism 
$\mathrm{exp}(\varepsilon \,\mathrm{pr}_3 V_2) \times \mathrm{id}  \colon J^3(\pi) \times \EuScript{V}\rightarrow J^3(\pi) \times \EuScript{V}$
maps the form $\tau_3$   to the Wahlquist--Estabrook form for the Lax representation
\begin{equation}
\left\{
\begin{array}{rcl}
\,v_t &=& [u,v]+u-\textfrac{1}{2}\,\mathrm{E}(u)+F(\mathrm{D}(u)+\varepsilon),      
\\
\,[\mathrm{D}(u), v] &=& \textfrac{1}{2} \,\mathrm{E}(\mathrm{D}(u))
+\gamma\, \mathrm{D}(u)+\gamma\,\varepsilon  
\end{array}
\right.
\label{third_covering_epsilon}
\end{equation} 
with the non-removable parameter $\varepsilon$.

\vskip 5 pt 
\noindent
{\sc Remark 1.}
The non-removability of the parameters in the differential coverings \eqref{third_covering_lambda}, \eqref{second_covering_epsilon}, and \eqref{third_covering_epsilon} can be proven directly, without reference to
\cite{KrasilshchikVinogradov1989,Krasilshchik2000,IgoninKrasilshchik2000,Marvan2002,IgoninKerstenKrasilshchik2002}.
To this end, we can check non-existence of the diffeomorphism $\Phi$ in \eqref{equivalence_of_coverings}. For example, to prove 
that the parameter $\lambda$ in \eqref{second_covering_epsilon} is non-removable, we consider the Wahlquist--Estabrook form 
$\tau_{v,\lambda} = \tau_2-(\lambda-1)\,u_{xyy}\,(1+ (v_x+\textfrac{1}{2}y)\,u_{xyy})^{-2}\,dy$ of this covering, the copy 
$\tau_{w,\mu}$ thereof, and the map $w=\Phi(t,x,y,u_I,v,v_x, ..., v_{mx})$ with some $m \ge 0$ and $\# I \le r$, 
where  $v_{kx} =\widetilde{D}_x^k (v_0)$,  $\widetilde{D}_x= D_x+\sum_{k \ge 0} v_{(k+1) x} \partial_{v_{kx}}$, and $v_0=v$.
Then the cumbersome induction with respect to $m$ and $r$   shows that the requirement for the form $\Phi^{*} \tau_{w, \mu}$ to belong to the ideal generated  by the forms $\widetilde{D}_x^k (\tau_{v,\lambda})$  and the restrictions of the contact forms  
$\vartheta_I\vert_{\EuScript{E}}$ is inconsistent  when $\mu \neq \lambda$, see \cite{KrasilshchikSergyeyevMorozov2016} for an example of such an induction in a similar proof.
\hfill $\diamond$

\subsection{The Lax representations for the Euler equation}

The inverse trans\-for\-ma\-ti\-on to the change of variables from \S 3.1 applied to systems  
\eqref{third_covering_lambda} and  \eqref{third_covering_epsilon} gives the following Lax representations for the Euler equation.
\vskip 5 pt 
\noindent
{\sc Theorem.}
{\it Systems 
\begin{equation}
\left\{
\begin{array}{rcl}
\,v_t &=& [u,v]+u-\textfrac{1}{2}\,\mathrm{E}(u)+\lambda\, F(\lambda^{-1}\, \Delta(u)),      
\\
\,[\Delta(u), v] &=& \textfrac{1}{2} \,\mathrm{E}(\Delta(u))
+\lambda\, G(\lambda^{-1}\, \Delta(u))  
\end{array}
\right.
\label{third_covering_lambda_for_Euler_equation}
\end{equation} 
with the function $G$ such that there holds \eqref{G_condition}   and
\begin{equation}
\left\{
\begin{array}{rcl}
\,v_t &=& [u,v]+u-\textfrac{1}{2}\,\mathrm{E}(u)+F(\Delta(u)+\varepsilon),      
\\
\,[\Delta(u), v] &=& \textfrac{1}{2} \,\mathrm{E}(\Delta(u))
+\gamma\, \Delta(u)+\gamma\,\varepsilon  
\end{array}
\right.
\label{third_covering_epsilon_for_Euler_eq}
\end{equation} 
define Lax representations for the Euler equation \eqref{Euler_eq}.  The parameters $\lambda$ and $\varepsilon$   in 
these systems are non-removable, that is, systems \eqref{third_covering_lambda_for_Euler_equation} 
and \eqref{third_covering_epsilon_for_Euler_eq} with  different constant values of their parameters are not equivalent. 
}
\hfill $\Box$

\vskip 5 pt 
\noindent
{\sc Remark 2.}
To obtain the Lax representation \eqref{Li_covering} we put $F \equiv 0$, $G \equiv 1$, and $v = \lambda\, w$ in 
 \eqref{third_covering_lambda_for_Euler_equation} . The resulting system   
\[
\left\{
\begin{array}{rcl}
\,
w_t &=& [u,w]+\frac{1}{\lambda}\,(u-\textfrac{1}{2}\,\mathrm{E}(u)),    
\\
\,[\Delta(u), w] &=& 1+\tfrac{1}{2\,\lambda} \,\mathrm{E}(\Delta(u)) 
\end{array}
\right.
\]
acquires  the form \eqref{Li_covering} as  $\lambda \rightarrow \infty$.
\hfill $\diamond$


\section{Concluding remarks}

We have shown that the technique of twisted extensions of Lie algebras combined with the transformation Lie--Rinehart algebras  
is useful for constructing new Lax representations of the 2D Euler equation, including the Lax representations with functional and  non-removable parameters.  We note that the results of the paper  provide a new example of twisted extensions of Lie algebras that are not of the Kac--Moody type.  We hope that our method will be applicable to other equations with  non-trivial second twisted cohomology groups of the symmetry algebras  that are  of importance in fluid dynamics, climate modeling, and magnetohydrodynamics. 
 Likewise, it is interesting to check whether the obtained Lax representations  can be implemented to study the Euler equation, in particular, to derive nonlocal symmetries and conservation laws thereof,  see 
\cite{MorozovSergyeyev2014,BaranKrasilshchikMorozovVojcak2016,KrasilshchikSergyeyevMorozov2016,%
BaranKrasilshchikMorozovVojcak2018,LelitoMorozov2018a,LelitoMorozov2018b,KrasilshchikMorozovVojcak2019,%
KrasilshchikVojcak2021} and references therein. We intend to address these issues in the  future work.


\section*{Acknowledgments}

I would like to express my sincere gratitude  to I.S. Kra\-{}sil${}^{\prime}$\-{}shchik for  very important discussions.
I thank M.V. Pavlov for useful remarks.

Computations of the generators of the symmetry algebra of equation \eqref{transformed_Euler_eq} were done using the {\sc Jets} software \cite{Jets}.

I thank the anonymous referee for several suggestions that helped improve the exposition of the paper.

\bibliographystyle{amsplain}

\begin{thebibliography}{10}



\bibitem{AblowitzClarkson1991} %
M.J. Ablowitz, P.A. Clarkson.  {\it Solitons, Nonlinear Evolution Equations and Inverse
Scattering}. Cambridge University Press, Cambridge, 1991

\bibitem{ArnoldKhesin} V.I. Arnold, B.A. Khesin. {\it Topological Methods in Hydrodynamics}. Springer, 1998

\bibitem{BaranKrasilshchikMorozovVojcak2016}
H. Baran, I.S.  Krasil$^\prime$shchik, O.I. Morozov, P. Voj\v{c}\'{a}k.
Coverings over Lax integrable equations and their nonlocal symmetries.
Theoretical and Mathematical Physics  {\bf 188} (2016), 1273--1295



\bibitem{BaranKrasilshchikMorozovVojcak2018}
H. Baran, I.S.  Krasil$^\prime$shchik, O.I. Morozov, P. Voj\v{c}\'{a}k.
Nonlocal symmetries of integrable linearly degenerate equations: a comparative study.
Theor. Math. Phys. {\bf 196 (2)} (2018),  1089--1110




\bibitem{Jets} H. Baran, M. Marvan. {\it Jets. A software for differential   calculus on jet spaces and diffieties}.
  
  {\tt http://jets.math.slu.cz}

\bibitem{Cheh_Olver_Pohjanpelto_2005}
J. Cheh, P.J. Olver, J. Pohjanpelto. Maurer--Cartan equations for Lie symmetry pseudo-groups of differential equations.
J. Math. Phys. \textbf{46} (2005), 023504

\bibitem{BacklundDarboux2001}
A. Coley, D. Levi, R. Milson, C. Rogers, P. Winternitz (eds).
{\it B{\"a}cklund and Darboux Trans\-for\-ma\-ti\-ons. The Geometry of Solitons}.
CRM Pro\-ce\-e\-dings and Lecture Notes, {\bf 28}, AMS, Providence, 2001

\bibitem{FelsOlver1998} 
M. Fels, P.J. Olver. Moving coframes. I.  A practical algorithm. Acta. Appl. Math. {\bf 51} (1998), 161--213 

\bibitem{Fuks1984} D.B. Fuks. {\it Cohomology of Infinite-Dimensional Lie Algebras}. Consultants Bureau, N.Y., 1986

\bibitem{IgoninKrasilshchik2000} %
S. Igonin, J. Krasil${}^{\prime}$shchik. On one-parametric families of B\"ack\-lund trans\-for\-ma\-ti\-ons.
In: T. Mo\-ri\-mo\-to, H. Sato, K. Yamaguchi (eds.),
{\it Lie Groups, Geometric Structures and Differential Equations --- One Hundred Years After Sophus Lie}. 
Advanced Studies in Pure Mathematics, {\bf 37}, pp. 99--114.
Math. Soc. Japan, Tokyo, 2002

\bibitem{IgoninKerstenKrasilshchik2002}
S. Igonin, P. Kersten, I. Krasil${}^{\prime}$shchik. 
On symmetries and cohomological invariants of equations possessing flat representations. 
 Diff. Geom. Appl. {\bf 19} (2003), 319--342

\bibitem{Konopelchenko1987} B.G. Konopelchenko. {\it Nonlinear Integrable Equations. Lecture Notes in Physics},
{\bf 270}, Sprin\-ger, 1987

\bibitem{Krasilshchik2000}
I.S. Krasil${}^{\prime}$shchik. On one-parametric families of B\"acklund transformations.
Preprint DIPS-1/2000, The Diffiety Institute, Pereslavl-Zalessky (2000)

\bibitem{KrasilshchikMorozovVojcak2019}
I.S. Krasil${}^{\prime}$shchik,   O.I. Morozov, P. Voj{\v{c}}{\'a}k. 
Nonlocal symmetries, con\-ser\-va\-tion laws, and recursion operators of the Veronese web equation.
Journal of Geometry and Physics {\bf 146} (2019), 103519

\bibitem{KrasilshchikSergyeyevMorozov2016}
I.S. Krasil$^\prime$shchik, A. Sergyeyev, O.I.  Morozov.
Infinitely many nonlocal conservation laws for the ABC equation with $A+B+C\neq 0$.
Calculus of variations and partial differential equations {\bf 55} (2016),
123, 12 pp.


\bibitem{KrasilshchikVinogradov1984} %
I.S. Krasil${}^{\prime}$shchik, A.M. Vinogradov. Nonlocal symmetries and the theory of coverings.
Acta Appl. Math. {\bf 2} (1984), 79--86

\bibitem{KrasilshchikVinogradov1989} %
I.S. Krasil${}^{\prime}$shchik, A.M. Vinogradov. Nonlocal trends in the geometry of differential equations:
sym\-met\-ri\-es, conservation laws, and B\"{a}cklund transformations.
Acta Appl. Math. {\bf 15} (1989), 161--209


\bibitem{KrasilshchikVojcak2021}
I.S. Krasil${}^{\prime}$shchik, P. Voj{\v{c}}{\'a}k. 
On the algebra of nonlocal symmetries for the 4D Martínez Alonso-Shabat equation.
Journal of Geometry and Physics {\bf 163} (2021), 104122


\bibitem{KruglikovMorozov2015}
B.S. Kruglikov, O.I. Morozov. 
Integrable dispersionless PDEs in 4D, their symmetry pseudogroups and deformations.
Lett. Math. Phys. {\bf 105} (2015),  1703--1723


\bibitem{LandauLifshits} 
L.D. Landau, E.M. Lifshitz. {\it Course of Theoretical Physics. Vol. 6. Fluid Mechanics}. 2${}^{\mathrm{nd}}$ English ed., revised.
Pergamon Press, Oxford, 1987


\bibitem{LelitoMorozov2018a}
A. Lelito, O.I Morozov.
Nonlocal symmetries of Pleba{\'n}ski’s second heavenly equation. Journal of Nonlinear Mathematical Physics {\bf 25:2} (2018), 188--197

\bibitem{LelitoMorozov2018b}
A. Lelito, O.I. Morozov, O.I. Three-component nonlocal conservation laws for
some integrable partial differential equations.
Journal of Geometry and Physics  {\bf 131} (2018), 89--100

\bibitem{Li2001}
Y.C. Li. A Lax pair for the two dimensional Euler equation. J. Math. Phys. {\bf 42} (2001), 3552--3553 

\bibitem{LiShvidkoy2004}
Y.C. Li, R. Shvidkoy. Isospectral theory of Euler equations. J. Math. Anal. Appl. {\bf 292} (2004), 311--315

\bibitem{LiYurov2003}
Y.C. Li, A.V. Yurov. Lax pairs and Darboux transformations for Euler equations. Stud. Appl. Math {\bf 111} (2003), 101--113

\bibitem{LouJiaHuangTang2007}
S.Y. Lou, M. Jia, F. Huang, X.Y. Tang. 
B\"{a}cklund transformations, solitary waves, conoid wa\-ves and Bessel waves of the (2+1)-dimensional Euler equation.
Int. J. Theor. Phys. {\bf 16} (2007), 2082--2095

\bibitem{LouJiaTangHuang2007}
S.Y. Lou, M. Jia, X.Y. Tang, F. Huang. Vortices, circumfluence, symmetry groups, and Dar\-boux transformations of the 
(2+1)-dimensional Euler equation. Phys. Rev. E {\bf 75} (2007), 056318 

\bibitem{LouLi2006} S.Y. Lou. Y.S. Li. Exact solutions of (2+1)-dimensional Euler equations found by weak Darboux transformation. 
Chin. Phys. Lett. {\bf 23} (2006), 2633--2636

\bibitem{Mackenzie1995}
K.C.H. Mackenzie. Lie algebroids and Lie pseudoalgebras. Bull. London Math. Soc. {\bf 27} (1995), 97--147


\bibitem{Marvan2002}
M. Marvan. On the horizontal gauge cohomology and nonremovability of the spectral parameter. 
Acta Appl. Math. {\bf 72} (2002), 51--65

\bibitem{MatveevSalle1991} V.B. Matveev, M.A. Salle. {\it Darboux Transformations and Solitons}.
Sprin\-ger, 1991

\bibitem{Morozov2002} 
O.I. Morozov. Moving coframes and symmetries of     differential equations. J. Phys. A, Math. Gen., {\bf 35} (2002), 2965--2977


\bibitem{Morozov2017}
O.I. Morozov. Deformed cohomologies of symmetry pseudo-groups and co\-ve\-rings of differential equations.
J. Geom. Phys., {\bf 113} (2017), 215--225

\bibitem{Morozov2018}
O.I. Morozov. Deformations of infinite-dimensional Lie algebras, exotic co\-ho\-mo\-lo\-gy, and integrable nonlinear partial differential equations. J. Geom. Phys., {\bf 128} (2018),  20--31

\bibitem{Morozov2019}
O.I. Morozov. Lax representations with non-removable parameters and integrable hierarchies of PDEs via exotic
cohomology of symmetry algebras. J. Geom. Phys., {\bf 143} (2019), 150--163

\bibitem{Morozov2021a}
O.I. Morozov. Nonlinear nonisospectral differential coverings for the hyper-CR equation of Ein\-stein--Weyl structures and the 
Gibbons--Tsarev equation.
Diff. Geom. Appl., {\bf 75} (2021),  101740

\bibitem{Morozov2021b}
O.I. Morozov.  Isospectral deformation of the reduced quasi-classical self-dual Yang--Mills equation. 
Diff. Geom. Appl., {\bf 76}  (2021),  101742

\bibitem{Morozov2022}
O.I. Morozov. Integrable partial differential equations and Lie--Rinehart algebras. J.Geom. Phys.,
{\bf 181} (2022),  104661 


\bibitem{Morozov2021c}
O.I. Morozov. Lax representations via twisted extensions of infinite-dimensional Lie algebras: some 
new results. "The Diverse World of PDEs: Geometry and Mathematical Physics". Edited by I. S. Krasil${}^{\prime}$shchik, 
A. B. Sossinsky, A. M. Verbovetsky.  Contemporary Mathematics, Vol. 788, AMS, 
Providence  RI, 2023, pp. 215--230




\bibitem{MorozovSergyeyev2014}
O.I. Morozov, A. Sergyeyev.
The four-dimensional Mart{\'{\i}}nez Alonso-Shabat equation: Reductions and nonlocal symmetries.
Journal of Geometry and Physics {\bf 85} (2014), 40--45

 
\bibitem{Novikov2002} S.P. Novikov. On exotic De-Rham cohomology. Perturbation theory as a spectral sequence.
    {\tt arXiv:math-ph/0201019}, 2002

\bibitem{Novikov2005} %
S.P. Novikov. On metric-independent exotic homology. Proc. Steklov Inst. Math. {\bf 251} (2005), 206--212

\bibitem{NovikovManakovPitaevskiyZakharov1984}
S.P. Novikov, S.V. Manakov, L.P. Pitaevskii, V.E. Zakharov. {Theory of Solitons}. Plenum Press, N.Y., 1984

\bibitem{Olver1993}
P.J. Olver. {\it Applications of Lie Groups to Differenial Equations}. 2${}^{\mathrm{nd}}$ Edition, Springer,
1993

\bibitem{Olver1995} 
P.J. Olver. {\it Equivalence, Invariants, and Symmetry}. Cambridge, Cambridge
    Uni\-ver\-si\-ty Press (1995)


\bibitem{Olver_Pohjanpelto_2005}
P.J. Olver, J. Pohjanpelto. Maurer-Cartan forms and the structure of Lie pseudo-groups. Selecta Math.
\textbf{11} (2005) 99--126

\bibitem{Olver_Pohjanpelto_Valiquette_2009}
P.J. Olver, J. Pohjanpelto, F. Valiquette. On the structure of Lie pseudo-groups. SIGMA \textbf{5} (2009), 077

\bibitem{RogersShadwick1982} C. Rogers, W.F. Shadwick. {\it B{\"a}cklund Transformations and Their Applications}.
Academic Pres, N.Y., 1982

\bibitem{VK1999} %
A.M. Vinogradov, I.S. Krasil${}^{\prime}$shchik (eds.) {\it Symmetries and Conservation Laws for Differential
Equations of Mathematical Physics} [in Russian],   Moscow: Factorial,  2005;
English transl. prev. ed.:
I.S. Krasil${}^{\prime}$shchik, A.M. Vinogradov (eds.) {\it Symmetries and Conservation Laws for Differential
Equations of Mathematical Physics}. Transl. Math. Monogr., {\bf 182}, Amer. Math. Soc., Providence, RI, 1999

\bibitem{WE} H.D. Wahlquist, F.B. Estabrook. Prolongation structures of nonlinear
    evo\-lu\-ti\-on  equations.  J. Math. Phys., {\bf 16} (1975), 1--7

\bibitem{YurovYurova2006}  A.V. Yurov, A.A. Yurova. One method for constructing exact solutions of equations of two-dimensional hydrodynamics of an incompressible fluid.  Theor. Math. Phys., {\bf 147:1} (2006), 501--508

\bibitem{Zakharov82} V.E. Zakharov. Integrable systems in multidimensional spaces.
    Lect. Notes Phys., {\bf 153} (1982), 190--216

\end{thebibliography}

\end{document}